# Classifying SMEs for Approaching Cybersecurity Competence and Awareness

Alireza Shojaifar
FHNW, IIT, 5210 Windisch, Switzerland
Utrecht University, Dept. of Information and Computing Sciences, Utrecht, Netherlands
alireza.shojaifar@fhnw.ch

Heini Järvinen
Tech.eu, Brussels, Belgium
heini@tech.eu

**ABSTRACT**

Cybersecurity is increasingly a concern for small and medium-sized enterprises (SMEs), and there exist many awareness training programs and tools for them. The literature mainly studies SMEs as a unitary type of company and provides one-size-fits-all recommendations and solutions. However, SMEs are not homogeneous. They are diverse with different vulnerabilities, cybersecurity needs, and competencies. Few studies considered such differences in standards and certificates for security tools adoption and cybersecurity tailoring for these SMEs.

This study proposes a classification framework with an outline of cybersecurity improvement needs for each class. The framework suggests five SME types based on their characteristics and specific security needs: cybersecurity abandoned SME, unskilled SME, expert-connected SME, capable SME, and cybersecurity provider SME. In addition to describing the five classes, the study explains the framework's usage in sampled SMEs. The framework proposes solutions for each class to approach cybersecurity awareness and competence more consistent with SME needs.

**KEYWORDS**

Cybersecurity awareness, Micro small and medium-sized enterprises, Classification, Capability improvement,

## 1 Introduction

Small and medium-sized enterprises (SMEs) are perceived to have the weakest defences against cyber-attacks [1, 2]. Many SMEs are often unaware of cybersecurity's significance, and the lack of adoption of precautions is a real risk [1, 3].

Diverse solutions proposed to provide training for awareness and cybersecurity capability improvement for SMEs. A vast amount of security advice is available [1]. ENISA developed training for raising awareness [4]. Other work described information security maturity assessments [5, 6], self-paced tools for training awareness improvement [e.g., 7, 8, 9], and information security management approaches [10, 11]. However, a report from Pnemon Institute [12] shows an increase in sophisticated cyber-attack against SMEs. A recent report from Hiscox [35] demonstrates a sharp increase in reported cyber-attacks among SMEs across UK, Europe, and the US. Many SMEs still lack awareness or do not adopt any of these solutions. One of the reasons for the lack of adoption may be that each of these approaches may fit some SMEs but not others.

SMEs are heterogeneous exhibit diverse cybersecurity needs, perceptions, and capabilities [13, 14]. For example, SMEs might have different Information System (IS) expertise, various cybersecurity self-efficacy, and diverse appreciation of cybersecurity threats [3, 18, 36, 37]. This diversity indicates that there is no one-size-fits-all. Consistency of security information with the target group's profile, including demographic factors, is imperative for delivering security content [1, 3, 15, 16, 17]. For example, the cybersecurity level of target audiences is vital to ensuring a cybersecurity program's success [15]. However, few studies have considered SMEs' differences and how to communicate and approach cybersecurity in a tailored manner [3, 18, 36]. Study [3] focuses only on cybersecurity standards and certification schemes adoption. Study [18] considers only two SME types, and [36] only studies individuals' risk perceptions and security management practices.

This study aims at addressing the heterogeneity problem with a classification framework. It distinguishes between categories of SMEs based on their characteristics. The characteristics include SME staff IT knowledge, cybersecurity offering, cybersecurity expertise in SME, awareness of threats and the importance of protection, awareness of good practices, and awareness of the dynamic essence of cybersecurity.

Classification framework is of vital importance since it reduces the complexity of approaching cybersecurity improvement by identifying security improvement needs for each class. The framework indicates that each type of SME needs a specific approach to be well protected. Therefore, instead of providing inefficient general recommendations and training content, cybersecurity communications effectively target each SME class. The study identified five types of SMEs, including cybersecurity provider, capable, expert-connected, unskilled, and abandoned SMEs. We argue that the classification can offer a significant contribution to the SME cybersecurity literature because SME classification has not yet been adequately served.

The remainder of this study is organised as follows. Section 2 presents the background of the research; section 3 outlines the classification framework for approaching cybersecurity improvement; section 4 explains the use of the framework in sampled SMEs; section 5 discusses the significance of the framework and future research avenues. Section 6 summarises and concludes.

## 2 Research Background

Classification is significant since it decreases the complexity of working with various entities with different features and reduces the amount of information we need to store [19, 20]. Defining concepts is important since "if we perceived each entity as unique, we would be overwhelmed by the sheer diversity of what we experience and unable to remember more than a minute fraction of what we encounter" [20]. Based on Smith and Medin [20], concepts allow us to go beyond the information given. When we assign an entity to a class on the basis of its perceptible attributes, we can infer some of its non-perceptible attributes. Category knowledge helps to make inferences about the presence of unobserved or unobservable features [19, 21].

Rosch [19] proposes two fundamental principles for classification: cognitive economy and perceived world structure. The cognitive economy refers to category systems' functions and indicates that category systems need to "provide maximum information with the least cognitive effort." Perceived world structure refers to the structure of the information so provided and indicates that "the perceived world comes as structured information rather than as arbitrary or unpredictable attributes." Therefore, "maximum information with least cognitive effort is achieved if categories map the perceived world structure as closely as possible."

Prior research considered SME classes in business and the characteristics in which SMEs differ widely from one another. Chua et al. [13] indicate that the characteristics of SME owner-managers, the aspects of the firm and its employees, and the characteristics of the environment in which they operate impact SME heterogeneity. Hagen et al. [22] provide evidence and introduce four distinct SME profiles and strategic business patterns.

Digital SME Alliance [3] highlights the importance of the analysis of different types of SMEs' cybersecurity requirements and consequently adapting the measures for effective cybersecurity adoption. Furthermore, the study based on Interim Report [23] confirms the impact of industry type and firm size on cybersecurity adoption. They identify four types of SMEs and their role in the digital ecosystem to tailor security standards:

- Digital enablers are SMEs that are active in developing and providing cybersecurity solutions.
- Digitally based are SMEs that cybersecurity is not the core of their business; however, they highly depend on digital and security solutions from the first category to ensure their business continuity.
- Digitally dependent are end-user SMEs that form the largest category of SMEs. They use regular ICT for running their businesses, and they need to access easily understandable and practical solutions.
- Start-ups are SMEs that security has a low priority since they are busy with the functional development of their business models. They need to understand the importance of security compliances and be motivated to adopt security standards.

Lee and Larsen [18] consider anti-malware software adoption in SMEs through a survey study. Their study indicates two types of SMEs (IT-intensive industries, non-IT intensive industries) and two types of SME executives (IS experts, non-IS experts). The study emphasises that vendor support, including the presence of designated technicians, easy access to technical assistance, $24 \times 7$ services, and periodic training, is a key facilitator in persuading executives to adopt security solutions. While the study explains the impact of industry type on adoption decision, it does not indicate a significant effect of the firm size on the adoption intention and actual adoption. Moreover, the study based on Protection Motivation Theory (PMT) [24] explains that SME executives' IS self-efficacy strongly influences cybersecurity adoption decisions.

Self-efficacy and outcome expectancies demonstrate individuals' perception of capabilities and capacities to perform specific required tasks successfully [25]. Self-efficacy is a motivational construct that influences individuals' initial choice of activities, goals, task engagement, and affective reactions to tasks. Moreover, it is a dynamic construct that can be changed due to learning, experience, and feedback [26].

Information system (IS) research has considered self-efficacy as a fundamental determinant of IS usage [27]. Organisational supports, including top management encouragement, impact employees' self-efficacy and IS usage [28]. Furthermore, since efficacy beliefs are situationally specific [27, 29], others have considered cybersecurity self-efficacy and used instruments to measure cybersecurity efficacy and skills [e.g., 30]. Competence in cybersecurity can be explained based on self-efficacy [30].

Classifying SMEs for Approaching Cybersecurity**Table 1: Improvement needs by SME cybersecurity class**

| SME Cybersecurity Classes | CSO | CSEA | ITE | CSTA | CSGP | CSAD | Cybersecurity Improvement Needs |
|---|---|---|---|---|---|---|---|
| Abandoned SMEs [a] | None | None | None | None | None | None | CS motivation, IT knowledge, CS knowledge, CS connection |
| Unskilled SMEs [b] | None | None | Yes | Partially | Partially | Adoption of CS Practices | CS training, CS guidance, CS connection |
| Expert-connect SMEs [c] | None | Internal/external CISO | Yes | Yes | Partially | Adoption of CS Practices | CS completion |
| Capable SMEs [d] | None | Expert | Yes | Yes | Yes | Continuous Improvement | CS news, CS evolution |
| Provider SMEs [e] | Yes | Expert | Yes | Yes | Yes | Continuous Improvement | CS news, CS evolution |

**CSO** = SME with Cybersecurity Offering; **CSEA** = staff, and CEO with cybersecurity expertise or in active contact with a cybersecurity expert; **ITE** = staff, and CEO with in-depth IT user Expertise; **CSTA** = staff and CEO with awareness about cyber threats and the importance of protection; **CSGP** = staff and CEO with awareness of SME-expected good cybersecurity practice; **CSAD** = CEO or Chief information security officer (CISO) aware about the dynamic character of cybersecurity

[a] Abandoned SMEs:
CS motivation: motivate the SME to adopt cybersecurity to overcome false beliefs about its true threat exposure,
IT knowledge: teach the SME's staff basic IT knowledge, including how to install, configure, and de-install software on devices,
CS knowledge: raise awareness about the most important cyber threats for the SME and recommendations for protection,
CS connection: connect the SME with a cybersecurity expert and peers that are improving their cybersecurity.

[b] Unskilled SMEs:
CS training: offer training to employees,
CS guidance: offer step-by-step instructions for implementing and maintaining SME-specific controls,
CS connection: connect the SME with a cybersecurity expert and peers that are improving their cybersecurity.

[c] Expert-connected SMEs:
CS completion: fill the gaps for protecting the SME given its business model.

[d, e] Capable SMEs and provider SMEs:
CS news: maintain awareness about incidents and changes in the threat landscape,
CS evolution: adapt the protection to changes in the threat landscape, CS and IT technologies, and the SME's business model.

Collective self-efficacy focuses on employees' aggregated capabilities instead of individual-focused and assessed by organisational representatives [18, 31]. SME executives or top managers are identified as individuals who can adequately assess their companies' collective self-efficacy. Also, their self-efficacy impact cybersecurity adoption in SMEs [18].

Bulgurcu et al. [30] indicate that providing organizational security awareness is an important factor in persuading employees to adopt security technologies and practices. They distinguish two types of awareness: general security awareness and information security policy (ISP) awareness. General security awareness is defined as an overall understanding of security threats, their consequences, and the importance of precautions. In addition, ISP awareness is defined as understanding the requirements prescribed in the policies and the aims of those requirements. Both types of awareness can be considered for SMEs.

Although the classification of SMEs is needed to tailor cybersecurity solutions, little attention has been given to it. Lee and Larsen [18] consider the importance of self-efficacy and expertise; however, categorising SMEs into IT-intensive and non-IT-intensive and the executives to expert and non-expert seems insufficient. DIGITAL SME [3] classifies SMEs to better adapt standards and certification schemes to the needs of SMEs in short to medium-term; however, the study explains that for the long-term goal, a mix of raising awareness and providing practical solutions is needed.

We now move to the classification framework to draw out approaching cybersecurity awareness-raising and capability improvement in various types of SMEs.

## 3  An SME Classification Framework for Approaching Cybersecurity Awareness

This section proposes a classification framework of five SME types and indicates cybersecurity improvement needs for each class (Table 1). The framework resulted from the paper design author experience with SMEs of six EU countries over several years on two projects. Iterative design security solutions for SMEs, using the design science research methodology [45], provided us the opportunity to learn more about SMEs and their differences. The concepts (classes) were defined to reflect maximum information about the SME characteristics and cybersecurity competence with the least cognitive effort to distinguish between the classes.

The following factors have been considered in the classification. The factors provide a minimal set, mutually independent to reflect competence and awareness in SMEs.

- SME with cybersecurity offering (CSO). The SME can be a cybersecurity provider company.
- Staff and CEO with cybersecurity expertise or in active contact with a cybersecurity expert (CSEA). The SME may have sufficient proficiency in cybersecurity or have internal/external CISO that support cybersecurity activities in the company or have no security expertise and connection to a security expert.
- Staff and CEO with in-depth IT user Expertise (ITE). The SME staff can have sufficient IT expertise or receive technical support from available resources.
- Staff and CEO with awareness about cyber threats and the importance of protection (CSTA). This factor reflects the SME staff's general perception of cybersecurity risks and the importance of implementing countermeasures.
- Staff and CEO with awareness of SME-expected good cybersecurity practice (CSGP). This factor reflects SME staff and CEO's understanding of the importance of guidelines and policies and the availability of a written policy in the company. The SME may have an explicit security guideline or policy statement in place according to the SME security requirements, or partially written for some focus areas, or no clear policy or guideline statement.
- CEO or CISO with awareness of the dynamic character of cybersecurity (CSAD). The SME approaches in cybersecurity can differ. If they realise that cybersecurity becomes obsolete, they may have a long-term attitude to plan updated training and review their policies. If they look at the awareness topics as secondary issues, they try to adopt security solutions to gain a security level. If they have no clear perception of potential threats and vulnerabilities, they are reluctant to adopt cybersecurity solutions.

According to the SME types, the approach of cybersecurity improvement needs to be adapted. Thus, the training awareness content or hands-on solutions would be more meaningful for SMEs. Five proposed classes are:

**Cybersecurity Abandoned SMEs**. In this type, SMEs have no cybersecurity policy or guideline. Along with a lack of security competence, IT skill shortages seem to constrain cybersecurity activities. They have no resource allocation or connection to cybersecurity resources. They have no clear perception of security threats; consequently, they do not see the need for security measures or commitment to cybersecurity practices. Providing extrinsic motivation to adopt security solutions and change incorrect beliefs about its true threat exposure is a significant need for this class.

Moreover, they need access to basic security and IT knowledge, hands-on skills, and training content to improve their capabilities. Further, connection to trusted security experts and peers for communication seems necessary. It can facilitate security controls implementation and knowledge transfer.

**Cybersecurity Unskilled SMEs**. In this type, SMEs have a partially written cybersecurity policy for some focus areas. They are aware of some security threats and vulnerabilities; however, they do not have a holistic view. They have a lack of cybersecurity skills. They are not connected to experts, third parties, or associations to exchange knowledge and develop their employees' skills, and therefore they lack the competence to manage cybersecurity measures. They realise the importance of cybersecurity measures and are willing to comply with security policies. Thus, access to hands-on security skills, training content, and cybersecurity experts can lead them to improve their capabilities and adopt security solutions.

**Cybersecurity Expert-connected SMEs**. This type of SME has a partially written policy for some focus areas. They are connected and dependent on trusted third parties or have a CISO to manage their cybersecurity measures. They are aware of the importance of cybersecurity, and they have a connection to gain knowledge and skills. The employees are not adequately skilled in cybersecurity; in turn, access to specific capabilities and training based on their business model can fill the cybersecurity gaps for protecting the SME.

**Cybersecurity Capable SMEs**. This type of SME has a cybersecurity culture and a written security policy fully aligned with what cybersecurity must be done, the same as the cybersecurity provider SMEs (the next class). However, the key differentiator between this type and security provider SMEs is their business model. They have expertise and proficiency in IT and cybersecurity. Access to the updated and newest cybersecurity and IT technologies material to adapt their protection approaches holds useful for this type. Also, access to cybersecurity news helps them maintain awareness about incidents and changes in the threat landscape.

**Cybersecurity Provider SMEs**. They provide security solutions for others. This type of SME has a cybersecurity culture and a written security policy the same as the capable SMEs (the previous class). They are aware that threats are ever-changing, so they regularly review their policy and updates their rules. Moreover, they have a plan to update their training for employees.

Thus, this type best demonstrates having a proactive attitude about cybersecurity activities. The same as cybersecurity-capable SMEs, their paramount cybersecurity need is access to the newest cybersecurity and IT technologies material and news (e.g., new policies, compromised websites).

## 4 The Use of the Framework

This section presents the early validation of the framework to provide evidence on the use and usefulness of the solution. The results are based on the first author qualitative study, interview, with five sampled SMEs (project partners). The participating SMEs have different sizes (micro, small, and medium) and are active in various industries. The selection of the subjects was based on their availability and their cybersecurity competence and experience level. This is an exemplar section to illustrate one example for each class of SME. This approach has been confirmed by [44].

**SME-1** is a micro-enterprise active in hair and beauty. The subject demonstrated no expertise in IT and cybersecurity. She was unaware of how a phishing attack can impact her business and her customers' data. Moreover, she did not know whom she should contact when an incident happens. Interestingly, she explained that:

"*I rank rather high my company security level.*"

Further, she did not indicate any specific security need. It seems she does not have a correct perception of cybersecurity threats.

According to the framework, the SME executive has the lowest level of self-efficacy; abandoned SME. Therefore, basic training for security awareness, cybersecurity motivation for implementing relevant security control, and supporting a connection to security and IT experts seem necessary.

**SME-2** is a small company active in the IT industry. The subject was willing to improve the SME's cybersecurity, and the company has a partially written policy for password management. However, the subject was unable to manage security measures and find relevant resources. The subject noted:

"*We do not have a security team department. If you do not have a CISO, [you need] offers [for] training classes and certification. We need delivery of services.*" Furthermore, the subject stated: "*We need to know how to solve the problems and not only presenting the problems.*"

According to the framework, it is an unskilled cybersecurity SME. Access to hands-on resources, training courses, and cybersecurity experts seems necessary.

**SME-3** is a micro health care company. The subject indicated specific training awareness requirements based on the company business model. He further explained that:

"*[general training content] is not applicable to us, the hardware that we use for the services is managed by third parties, and they also set up the network. We need training content about cloud services for training the employees.*"

According to the framework, it is a cybersecurity expert-connected SME. A third party is responsible for managing their cybersecurity measures. Although the SME staff are aware of potential security threats, they do not have enough cybersecurity competence according to their business model. Access to specific training awareness content congruent with their business model seems useful.

**SME-4** is a medium-sized company active in electronic voting technologies. The company has a security department as well as a written policy. The subject noted that:

"*Access to material for more advanced security controls such as trusted boot or hardware encryption or a list of the latest threats and vulnerabilities is useful [for us].*"

According to the framework, it is a cybersecurity capable SME. Access to the latest updates and advanced security controls seems useful.

**SME-5** is a small company active in cybersecurity. The company provides security solutions and advice to other firms. The company has a written policy in place and puts a high value on review and update security measures. The subject indicated that:

"*We review our policy two or three times a year. Having the most recent updates and news are useful to review.*"

According to the framework, it is a cybersecurity provider SME. The company staff has great cybersecurity competence, and the same as the cybersecurity-capable SMEs, access to the latest updates in cybersecurity and IT seems useful.

## 5 Discussion

The contribution of this study is proposing an SME classification framework and indicating cybersecurity improvement needs for each SME type. The framework can reduce the complexity of SME heterogeneity and the lack of security adoption, leading to targeting more effective cybersecurity competence and awareness.

Commonly studies distinguish between SMEs based on the number of employees [37, 38, 39]. However, it is not enough to approach effective cybersecurity in SMEs. In line with [3, 18], this study demonstrates that classification helps enrich the understanding of SME types to communicate and keep them engaged effectively. The classification approach is in contrast with the idea of CYSFAM [43] that proposes a maturity model for generic organisations. Moreover, compared to [3, 18] (which identify four and two types of SMEs, respectively), this study indicates five types of SMEs with no counterpart for the cybersecurity capable SMEs.

The proposed framework is not a maturity model, and it does not convey that one class is more secure or vulnerable than the others. Instead, it is a taxonomy of distinct SME types and indicates that each type exhibits different needs to be secured. Therefore, the framework does not signify that there is a progression from one class to another one. While there are predictable reasons for movement between classes, there is no necessary sequence between the SME classes. For instance, an unskilled SME can hire an internal CISO, or an abandoned SME may establish a connection to a security provider SME and, consequently, move to the expert-connected class.

SMEs are heavily restricted with the available funding for cybersecurity purposes [41]; however, cybersecurity projects and service providers are approaching security in SMEs by developing cost-effective and lightweight solutions. The framework can help these service providers understand the level of cybersecurity expertise and good practices of SMEs in the different categories. Even more importantly, it shows the need to reach out to the potential end-users of their solutions with a messaging that focuses on the improvement needs of each category. The improvement needs of each category vary greatly, and there is little overlap. This means that a cybersecurity service provider must choose between the target audiences or markets it prioritises when it comes to communications, messaging, and even offering services and tools. For example, the European Horizon 2020 project GEIGER [42] could specialise first in one of the categories and focus on capturing its interest with the communications highlighting its specific improvement needs, and then extend the services and communications to reach the rest of the groups.

In the context of GEIGER, the key contents of communications targeted to the different categories could be:

- Abandoned SMEs: raising awareness of the existence and importance of addressing cybersecurity threats, teaching basic IT skills and how to evaluate risks, recommendations, and connecting with experts and tool providers.
- Unskilled SMEs: offering beginner or intermediate level training packages and connecting with experts ("Digital Security Defenders") who can provide concrete support in implementing the given recommendations.
- Expert-connected SMEs: connecting with experts who can assist in detecting the remaining weak areas and in establishing robust good practices for daily operations and continuous improvement.
- Capable SMEs and provider SMEs: highlighting the features of the offered tool that allow for continuous monitoring of and adaptation to the threat landscape and novel tools and technologies.

To raise the chosen target audience's interest and convince them, messaging highlighting their improvement needs should be consistently implemented throughout all channels. Consistent security messages affect SMEs' threat appraisal [1] and motivate them to implement necessary but straightforward precautions [1, 18, 32]. For example, if choosing to focus on the abandoned SMEs category, the essential contents of the landing page of the GEIGER solution could include a catchy and concrete story of a peer SME who discovered their cybersecurity risks and started improving them with the help of GEIGER. Also, a short questionnaire to evaluate their current risks. CEOs in abandoned SMEs may have incorrect perceptions of their security level and potential risks. So, they might be demotivated to adopt cybersecurity solutions. Julisch argues that SMEs may argue "nobody would want to attack us" [40]. Beliefs and perceptions affect users' intention of cybersecurity activities [30]. Furthermore, GEIGER could support abandoned SMEs' IT skills. They lack the technical IT expertise, affecting the GEIGER solution adoption. The lack of IT and computer self-efficacy impacts security solution adoption [33, 34], and in SMEs is a significant inhibitor [18].

As the businesses in the categories of abandoned and unskilled SMEs have low awareness of cybersecurity issues, it is likely to be most efficient to reach out to them through non-cybersecurity-related channels that they already follow for professional or personal purposes. For example, trade or association newsletters and publications or presence at industry events, as well as direct contacts through their trusted service providers (such as accountants) or peer SMEs. The three other categories could, in addition, be reached through channels and events linked to cybersecurity.

This solution paper proposed a framework and exemplar section to apply it based on one sampled SME for each category. The avenue for future research is to empirically validate the framework across a broader sample of SMEs using, for instance, a survey-based quantitative approach studying the diversity of the SMEs in categories and elaborate their needs in more detail. Further, future work needs to entail more metrics for SME classification, for instance, concerning privacy needs, if SMEs that need to process personal information have active contact with Data Protection Officer (DPO). However, this study takes its place among the very few studies in SMEs' classification for cybersecurity improvement.

## 6  Conclusion

The paper has proposed a classification framework to better target value-ridden cybersecurity improvement in various types of SMEs. Based on SME characteristics, the framework identified five SME types: cybersecurity abandoned SME, unskilled SME, expert-connected SME, capable SME, and provider SME. Moreover, the framework studied different cybersecurity needs for approaching security improvement in each class.

Further, the study illustrated the use of the framework in the sampled SMEs from different industries. The early validation of the framework demonstrated that the framework could explain the differences between SME types. Moreover, the subjects identified some needs that have been considered in the framework. The security needs constituted a broad diversity. Cybersecurity unskilled and abandoned SMEs needed to connect to security experts and access training awareness material. The expert-connected SME mainly required capabilities to fill specific security gaps, and capable and provider SMEs needed to have updated and newest cybersecurity and IT technologies material.

The framework aims to demonstrate how each class of SME can be effectively communicated and well protected and does not convey that one class is more secure or vulnerable than the others. Therefore, the framework can help cybersecurity service providers in that they can position SMEs in one of the classes in the early face to decide how to communicate and offer services and tools.

## ACKNOWLEDGMENTS

This work was made possible with funding from the European Union's Horizon 2020 research and innovation programme, under grant agreement No. 883588 (GEIGER). The opinions expressed and arguments employed herein do not necessarily reflect the official views of the funding body.